# OPTICAL DETERMINATION AND IDENTIFICATION OF ORGANIC SHELLS AROUND NANOPARTICLES: APPLICATION TO SILVER NANOPARTICLES

T. MAURER, N. ABDELLAOUI, A. GWIAZDA, P.-M. ADAM, A. VIAL and J.-L BIJEON

*Laboratoire de Nanotechnologie et d'Instrumentation Optique, ICD CNRS UMR STMR 6279, Université de Technologie de Troyes ,12 rue Marie Curie, CS 42060, 10004 Troyes Cedex*
*thomas.maurer@utt.fr*

D. CHAUMONT and M. BOUREZZOU

*Nanoform - Laboratoire Interdisciplinaire Carnot de Bourgogne, ICB CNRS UMR 6303, Université de Bourgogne, BP 47 870, 21 078 Dijon cedex, France*
*denis.chaumont@u-bourgogne.fr*

**We present a simple method to prove the presence of an organic shell around silver nanoparticles. This method is based on the comparison between optical extinction measurements of isolated nanoparticles and Mie calculations predicting the expected wavelength of the Localized Surface Plasmon Resonance of the nanoparticles with and without the presence of an organic layer. This method was applied to silver nanoparticles which seemed to be well protected from oxidation. Further experimental characterization via Surface Enhanced Raman Spectroscopy (SERS) measurements allowed to identify this protective shell as ethylene glycol. Combining LSPR and SERS measurements could thus give proof of both presence and identification for other plasmonic nanoparticles surrounded by organic shells.**

The synthesis of plasmonic nanoparticles (NPs) via top-down processes like Electron Beam Lithography [1, 2], block copolymer photolithography [3], or nanosphere lithography [4, 5] but also via bottom-up ways like chemical routes [6] have been deeply studied due to their high potential for applications in medicine [7], biosensing [8, 9], hybrid photonic circuits [10, 11], optical devices [12, 13],… Among the family of plasmonic nanoparticles, silver nanoparticles exhibit appealing features like giant Surface Enhanced Raman Spectroscopy (SERS) enhancement [14, 15]. In spite of their unique properties, Ag nanoparticles appear to be less attractive for applications than gold NPs whose properties are not altered by oxidation and thus are more stable [16]. In this paper, we present a chemical route which easily provides Ag nanoparticles stable in air. The synthesis is based on a polyol process [17] assisted with microwave heating [18] which guarantees well controlled size dispersions. We argue here that an ethylene glycol shell at the surface of the Ag NPs prevents them from oxidation. In order to confirm the hypothesis, we made optical measurements and compared them to Mie simulations with or without the presence of ethylene glycol. We show in this paper that simple optical measurements compared to numerical simulations are a very useful tool to determine the presence of organic shell at the surface of plasmonic NP. Eventually, the identification of the protective organic shell was provided by SERS measurements. The advantages of combining such optical characterization techniques will be discussed in conclusion.

The Ag NPs are synthesized via a polyol process assisted with microwave heating [17,18]. $AgNO_3$ salts ($[AgNO_3]$ = 7.8 $10^{-3}$ mol/L) and Poly Vinyl Pyrrolidone ($[PVP]$ = 1.3 $10^{-3}$ mol/L) were dissolved in 6mL of ethylene glycol. The Ag NPs were synthesized after 20min heating by a microwave-assisted reflux with a 600W power. The microwave-assisted reflux synthesis of nanoparticles is well-adapted since it improves the reaction kinetic and homogeneity. After micro-wave heating, the solutions were decanting before washing. The washing step consisted in separating the Ag NPs from the solution by centrifugation at 4500rpm speed during 20min after dilution in 200mL of ethanol. This washing was applied 6 times. Then the Ag NPs were dried in an oven at 60°C for at least 12 hours. The obtained NPs are more or less polyhedral (Figure 1). The SEM images analysis of a hundred of NPs allows concluding that the average diameter of the size distribution is 43nm and the standard deviation 15%.

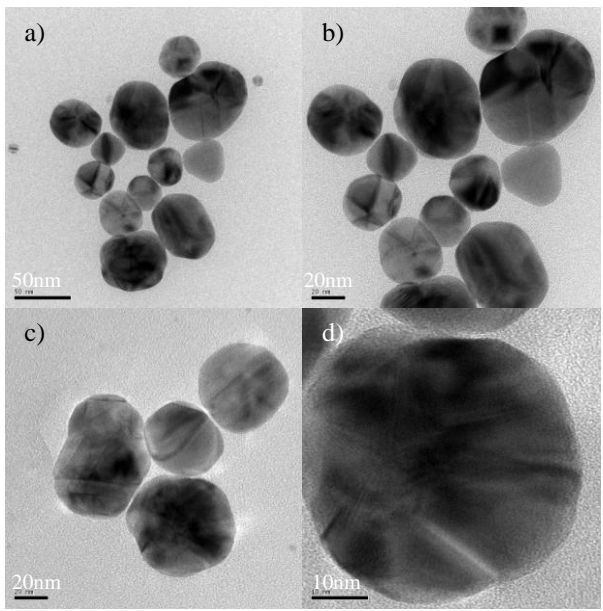

Fig 1. TEM images of Ag nanoparticles after drying for 12h. The size of the NPs is ranked between 15nm and 50nm

X-ray Diffraction (XRD) measurements were then performed on dried powder one week after its collection (see Figures 2a and 2b). These measurements are in good agreement with the ICDD Powder Diffraction file n°03-065-2871. The nanoparticles exhibit a face-centered cubic structure and the Scherrer formula, calculated for each peak, provides an average crystallite size of 12.2nm which indicates that the structure of the Ag nanoparticles is polycrystalline. XRD measurements were also performed on the same sample three years later (see Figures 2c and 2d). No evolution can be seen on the XRD pattern so that no oxidation or particle fragmentation can be assumed. Moreover, Energy-Dispersive X-ray (EDX) spectroscopy was performed to get access to the chemical composition of the nanoparticles. It showed that the nanoparticles which were synthesized are indeed mainly made of silver since its atomic pourcentage is about 92.6%. The other 7.4% corresponds to the presence of oxygen which could come either from air pollution during the measurements or from the presence of a polymer layer at the nanoparticle surface. Thus, it confirms that the Ag NPs are protected from full oxidation. At this stage, the question becomes to understand what prevents the Ag NPs from oxidation. Our hypothesis is the presence of an organic shell –and in particular an ethylene glycol layer- at the surface of the Ag NPs [19]. This would explain the small percentage of oxygen found in the EDX measurements but also the stability of the XRD measurements even 3 years later.

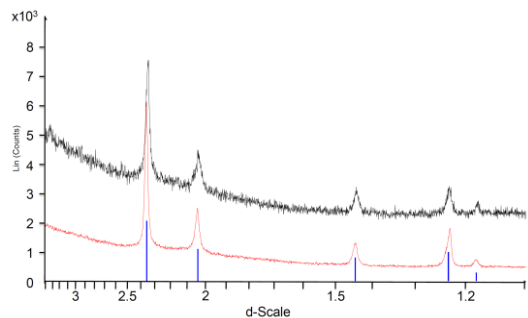

Fig. 2. black curve) XRD measurements after one week, red curve) XRD measurements after 3 years. The vertical lines correspond to the Ag° scattering peaks from the ICDD Powder Diffraction file n°03-065-2871. These results show that the NPs remain stable in air since no evolution of the XRD pattern is observed.

To prove this hypothesis, we dispersed the Ag NPs on glass substrates coated with a 30nm layer of ITO (Indium Tin Oxide) in order to perform optical extinction measurements on isolated Ag NPs (see Figure 3a) and to compare the Localized Surface Plasmon Resonance (LSPR) mode of these nanoparticles (see Figure 3b) to the one obtained via Mie calculations on Ag NPs with or without organic shell.

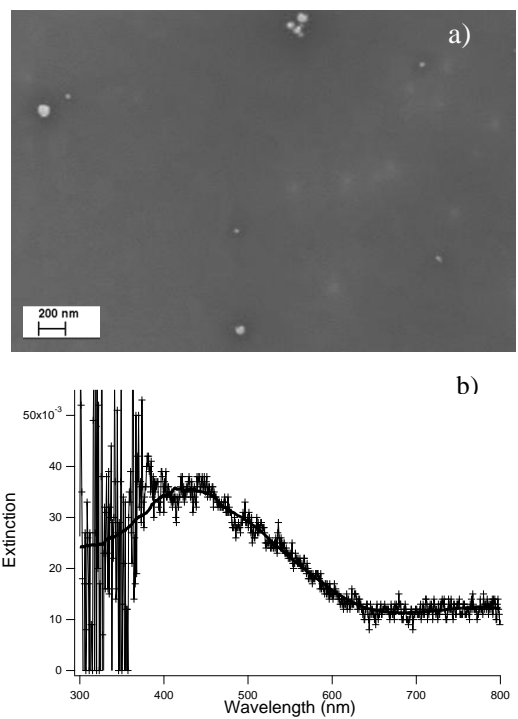

Fig. 3. a) SEM image of Ag NPs which have been highly diluted in ethanol solution before spin-coating on an ITO(30nm)/glass substrate. It can be observed that the NPs are isolated, b) optical extinction measurements of these Ag Nps over a 5µm large area. The wavelength of the LSPR peak is 420nm.

Then we performed a set of simulations using the Mie theory [20]. We considered core-shell spherical structures [21], with a silver core of fixed radius, and a silica shell with variable thickness which plays the role of the ethylene glycol organic shell because of the similar optical index. In Figure 4, we present the extinction efficiency for a core of diameter 80 nm, with a silica shell of thickness from 0 to 40 nm. The shift of the dipolar resonance from 380 nm up to 460 nm can be clearly observed when the silica thickness is increased. A weaker quadrupolar resonance can also be seen at a lower wavelength. Compared to Figure 3b, it tends to prove that there is the presence of shell around the Ag core. Indeed, if there was no shell, the LSPR wavelength would be below 400nm as indicated in Figure 4.

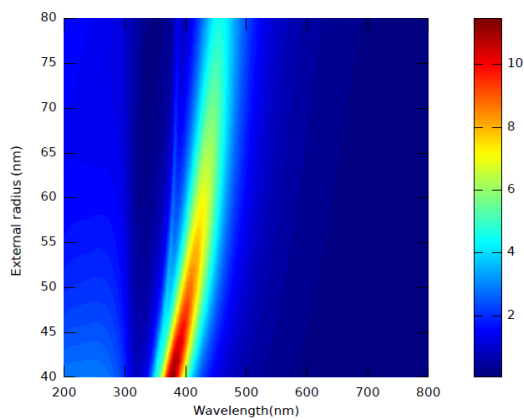

*Fig 4. Extinction efficiency for a core of diameter 80 nm with a silica shell of thickness 0 to 40 nm simulating using Mie theory[20,21].*

Besides, in order to get direct evidence of the presence of ethylene glycol, Surface Enhanced Raman Spectroscopy of Ag NPs was performed (see Fig. 5). Raman scattering is a powerful tool to identify chemical compounds since it provides a cartography of their chemical bounds and thus allows to deduce their nature. The inelastic behavior of processes involved in Raman scattering does not allow detecting small quantities of molecules. However, it has been demonstrated many times that plasmonic nanoparticles (in particular silver ones) can so strongly enhance the Raman scattering signal that single molecule may be detected [22]. In the framework of our study, silver nanoparticles can therefore play the role of SERS substrates to allow detecting their own organic shell and thus confirming the LSPR/numerical results. Each molecule has a unique Raman scattering signature which can be found in handbooks or publications. It has been previously reported that the Raman spectra of solid and liquid ethylene glycol may slightly differ since in solid state, ethylene glycol only exists in the gauche form [23]. Nevertheless, we measured the Raman spectrum of liquid ethylene glycol and compared it to the one of the Ag NPs in order to see if common peaks could be observed (see Figure 5). Moreover, the comparison with classical Raman assignments for solid ethylene glycol [23] indisputably confirms that the Ag NPs are covered and thus protected by an ethylene glycol layer (see Figure 5).

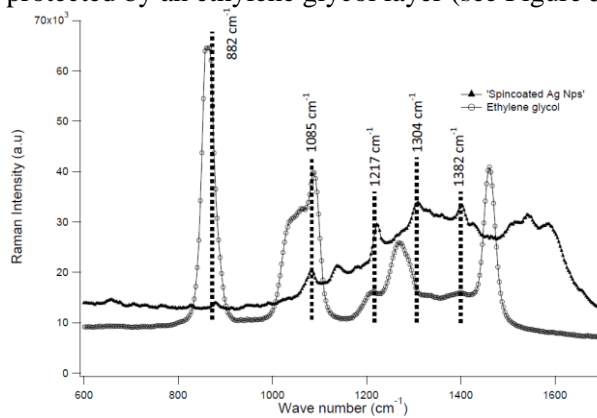

*Fig. 5. Raman spectrum of spincoated (triangle markers) Ag NPs compared to Raman spectrum of liquid ethylene glycol. The peaks related to solid ethylene glycol are indicated by the dashed lines [23].*

Finally, it appears that the polyol process assisted by microwave heating is a very appealing method for synthesis Ag nanoparticles since it provides large quantities of Ag nanoparticles protected from oxidation. It is usually admitted that despite of less appealing plasmonic features in the visible region of the spectrum- silver particles displaying sharper resonances- gold nanoparticles exhibit higher potential for applications due to their stability in time. This is why the silver nanoparticles presented here deserve much attention for application development, in particular as transducers in optical sensors [24,25]. Moreover, the second striking result of this paper consists in comparing optical extinction measurements and Mie simulations to prove the presence of organic layer. This very simple approach can be extended to other systems for which the presence of a shell will significantly red-shift the LSPR mode [19]. It prevents from using techniques which are not necessarily available in all laboratories such as Small-Angle X-ray Scattering (SAXS) or Small-Angle Neutron Scattering (SANS) [26]. Moreover, the use of SERS measurements is crucial since it may help to the identification of this organic shell which is not always possible via SAXS and SANS.

**Acknowledgments**

Financial support of NanoMat (www.nanomat.eu) by the "Ministère de l'enseignement supérieur et de la recherche," the "Conseil régional Champagne-Ardenne," the "Fonds Européen de Développement Régional (FEDER) fund," and the "Conseil général de l'Aube" is acknowledged. T. M thanks the CNRS for financial support via the « optical nanosensors » chaire.